\documentclass[aps,prd,reprint,showpacs]{revtex4-1} 
\usepackage{amsmath,graphicx} 

\newcommand*\al{\alpha}
\newcommand*\be{\beta}
\newcommand*\ga{\gamma}
\newcommand*\de{\delta} 

\newcommand*\ve{\varepsilon}

\newcommand*\et{\eta}

\newcommand*\la{\lambda}

\newcommand*\si{\sigma}

\newcommand*\ch{\chi}
\newcommand*\ps{\psi}

\newcommand*\mn{{\mu\nu}} 

\newcommand*\prt{\partial}

\newcommand*\ket[1]{|{#1}\rangle}
\newcommand*\fr[2]{{\textstyle{{#1} \over {#2}}}}
\newcommand*\half{{\textstyle{1\over 2}}}
\newcommand*\lsim{\mathrel{\rlap{\lower4pt\hbox{\hskip1pt$\sim$}}
    \raise1pt\hbox{$<$}}}
\newcommand*\gsim{\mathrel{\rlap{\lower4pt\hbox{\hskip1pt$\sim$}}
    \raise1pt\hbox{$>$}}}
\newcommand*\sqr[2]{{\vcenter{\vbox{\hrule height.#2pt
         \hbox{\vrule width.#2pt height#1pt \kern#1pt
         \vrule width.#2pt}
         \hrule height.#2pt}}}}

\newcommand*\abs[1]{\left|{#1}\right|}

\newcommand*\tb{\tilde{b}} 
\newcommand*\td{\tilde{d}} 
 
\newcommand*\tc{\tilde{c}} 
\newcommand*\tH{\tilde{H}} 

\newcommand*\citelabel[1]{ \label{#1} } 
\newcommand*\labeleeq[1]{ \label{#1} \end{equation} } 
\newcommand*\labeleea[1]{ \label{#1} \end{eqnarray} }

\newcommand{\beq}{\begin{equation}}
\newcommand{\eeq}{\end{equation}}
\newcommand{\bea}{\begin{eqnarray}}
\newcommand{\eea}{\end{eqnarray}}

\newcommand{\eqs}[1]{Eqs.\ (\ref{#1})}

\newcommand{\eq}[1]{Eq.\ (\ref{#1})}
\newcommand{\Eq}[1]{Eq.\ (\ref{#1})}

\newcommand{\citemath}[1]{\text{\cite{#1}}} 

\begin{document}

\preprint{Berry College HEP-004} 

\title{Spacetime Variation of Lorentz-Violation Coefficients at Nonrelativistic Scale} 

\author{Charles D.\ Lane}
 \email{clane@berry.edu}
\affiliation{Physics Department, Berry College, Mount Berry, GA 30149-5004} 
\affiliation{Indiana University Center for Spacetime Symmetries, Indiana University, 
 Bloomington, Indiana 47405-7105} 

\date{\today}

\begin{abstract} 
The notion of uniform and/or constant tensor fields of rank $>0$ 
is incompatible with general curved spacetimes. 
This work considers the consequences of certain tensor-valued coefficients for Lorentz violation 
in the Standard-Model Extension varying with spacetime position. 
We focus on two of the coefficients, $a_\mu$ and $b_\mu$, that characterize Lorentz violation 
in massive fermions, 
particularly in those fermions that constitute ordinary matter. 
We calculate the nonrelativistic hamiltonian describing these effects, 
and use it to extract the sensitivity of several precision experiments 
to coefficient variation. 
\end{abstract} 

\maketitle


\section{Introduction}

Local Lorentz symmetry is known to hold to a very high degree in our Universe 
\cite{Kostelecky:2008ts,Will:2005va}. 
However, there remains the possibility that it is broken 
and that this breaking might manifest itself in extremely precise experiments. 

This work adopts the minimal Standard-Model Extension (SME) in curved spacetime 
\cite{Kostelecky:1994rn,Colladay:1996iz,Colladay:1998fq,Kostelecky:2003fs,Kostelecky:2013rta} 
as a general framework for describing violations of particle Lorentz symmetry. 
This framework has been used to study many high-precision tests of Lorentz violation, 
including those that probe interactions involving the constituents 
of ordinary matter: electrons, neutrons, and protons 
\cite{Kostelecky:2008ts,Kostelecky:1999mr,Heckel:2008hw,Kornack:2008zz,Hohensee:2013cya, 
Altschul:2010na,Altschul:2006pv,Stadnik:2014xja,Peck:2012pt,Brown:2010dt,Wolf:2006uu, 
Allmendinger:2013eya,Cane:2003wp}. 

Lorentz violation in the sector of the minimal SME \cite{Kostelecky:2013rta} 
describing ordinary matter 
is parameterized by the set of coefficients $a_\mu, \ldots, H_\mn$. 
Most previous comparisons \cite{Kostelecky:2008in,Kostelecky:2010ze,Seifert:2010uu} 
between the SME and experimental results 
assume that these coefficients do not vary with spacetime position. 
That is, they assume that $\prt_\al a_\mu \equiv \cdots \equiv \prt_\al H_\mn \equiv 0$. 
In curved spacetime, however, this assumption cannot pertain. 
Statements involving partial derivatives such as $\prt_\al a_\mu\equiv 0$ are coordinate dependent, 
and therefore may hold in a only a limited set of special frames. 
There is no {\itshape a priori} reason for any experimental frame 
to be one of these special frames. 

One may instead try to impose the coordinate-independent condition $D_\al a_\mu \equiv 0$, 
but this assumption is generally incompatible with nonzero curvature \cite{Kostelecky:2003fs}. 
It implies that ${R^\mu}_{\al\nu\be}a_\mu \equiv 0$, 
which can only occur if the spacetime has at least one flat direction 
and if $a_\mu$ points along that direction 
\footnote{One simple example is $\mathbf{R}\times S^2$. 
Its Riemann tensor is nonzero, 
but nonzero vector field $a_\mu=(1,0,0)$ satisfies $D_\al a_\mu\equiv 0$. }. 
The spacetimes that are relevant for comparison to experiment, 
such as Schwarzschild spacetime, 
do not satisfy this requirement. 
In this work, 
we assume that $\prt_\al a_\mu \ne 0, \ldots, \prt_\al H_\mn\ne 0$
and $D_\al a_\mu \ne 0, \ldots, D_\al H_\mn\ne 0$ in general. 

This article is organized as follows. 
Section \ref{Basics} collects several small preliminary discussions: 
conventions, 
fundamental framework, 
and some rough estimation of effect sizes. 
Section \ref{explicit} gives a full expression 
for the relevant nonrelativistic hamiltonian. 
In Section \ref{analysis}, 
we isolate the dominant terms, 
study the C, P, and T properties of derivative interactions, 
and list the sensitivity of already-completed experiments 
to derivatives of SME coefficients. 
A summary appears in Sec.\ \ref{Summary}. 

\section{Basics}
\citelabel{Basics} 

\subsection{Conventions and Framework} 
\citelabel{conventions} 

We use Greek indices to denote spacetime coordinates $0,1,2,3$. 
Latin indices from the beginning of the alphabet $\{a,b,c\}$ denote local Lorentz coordinates $0,1,2,3$, 
while Latin indices near the middle of the alphabet $\{j,k,\ldots, q\}$ denote local spatial coordinates $1,2,3$. 
We work in a spacetime of metric signature $+2$, 
so that the flat-spacetime metric $\et_{ab}$ is $\text{diag}(-1,+1,+1,+1)$ 
and $p^j=-i\prt^j$ is the free-particle momentum operator. 
For ease of application to nonrelativistic systems, 
we work in the Dirac representation of the gamma matrices: 
$\ga^0=\left( \begin{array}{cc} \openone & 0 \\ 0 & \openone \end{array} \right)$ and 
$\ga^j=\left( \begin{array}{cc} 0 & \si^j \\ -\si^j & 0 \end{array} \right)$, 
where $\openone$ is the $2\times 2$ identity matrix and 
the usual Pauli matrices are denoted by $\si^j$. 
We define the Levi-Cevita symbol so that 
$[\si^j,\si^k]=2i{\ve^{jk}}_{l}\si^l$, 
which corresponds to the choice ${\ve^{12}}_{3}=+1$. 
We use the shorthand notation $\hbar_{jk}:=h_{jk}+\et_{jk}h_{00}$ 
for a combination of components of metric perturbation $h_\mn$ that appears often. 
Symmetrization/antisymmetrization involving parentheses/brackets around a pair of indices 
includes a factor of $1/2$: 
$T^{(jk)}:=\half(T^{jk}+T^{kj})$ 
and $T^{[jk]}:=\half(T^{jk}-T^{kj})$. 

Lorentz symmetry is known to hold to a very high degree in our universe, 
and therefore we can expect coefficients for Lorentz violation to be very small. 
We thus keep only terms up to first order in Lorentz-violation coefficients 
throughout this work. 

Our fundamental framework is 
the minimal Standard-Model Extension (SME) for a free Dirac fermion 
in weakly curved spacetime with no torsion. 
Specifically, 
we work in a spacetime frame where the background metric may be written 
$g_\mn = \et_\mn + h_\mn$ with each component $\abs{h_\mn}\ll 1$. 
With this assumption, 
we neglect any effects that are higher than first order in $h_\mn$. 
Further, 
we restrict attention to the SME coefficients $a_{\mu}$ and $b_{\mu}$, 
assuming that all others $c_{\mu\nu}, \ldots, H_{\mu\nu}$ are identically zero. 
This corresponds to the action \cite{Kostelecky:2003fs} 
\begin{align} 
\label{action} 
S_\ps = \int d^4 x & \left\{ 
 \half i 
 \left( \de^\mu_a -\half h^\mu_a +\half \de^\mu_a h^\al_\al \right) 
 \left( \overline{\ps}\ga^a\prt_\mu\ps - \prt_\mu\overline{\ps}\ga^a\ps \right) 
 \right. \nonumber \\ 
& \left. 
 -\overline{\ps} \left[ 
  (1+\half h^\al_\al)m 
  +\fr{i}{16} \prt_c h_{ab} \{\ga^a,[\ga^b,\ga^c]\} 
 \right] \ps 
 \right. \nonumber \\ 
& \left. 
 -\overline{\ps} 
 \left( \de^\mu_a -\half h^\mu_a +\half h^\al_\al \de^\mu_a \right) 
 \left( a_\mu\ga^a + b_\mu\ga_5\ga^a \right) 
 \ps 
 \right\} \quad . 
\end{align} 
The first two lines in $S_\ps$ are just the usual Lorentz-invariant action 
for a Dirac fermion in weakly curved spacetime, 
correct to first order in $h_\mn$. 
The third line contains the coefficient fields for Lorentz violation $a_\mu$ and $b_\mu$, 
which may depend on spacetime position. 

After performing a field redefinition $\ps=A\chi$ \cite{Colladay:1996iz,Lehnert:2004ri} 
to ensure conventional time evolution, 
applying Euler-Lagrange equations, 
and solving for $i\prt_0 \ch = H\ch$, 
we find a relativistic $4\times 4$ hamiltonian $H$. 
This appears identical to the hamiltonian found in \cite{Kostelecky:2010ze}, 
though in the present work we consider $a_\mu$ and $b_\mu$ to depend on spacetime position. 
It can be organized as 
\beq 
\label{Hrel} 
H = m\ga^0 + \mathcal{P} + {\cal E} + {\cal O} \quad , 
\eeq 
where 
\bea 
\label{PEO} 
{\cal P} &=& \ga^0 \ga^j p_j \quad , \nonumber \\ 
{\cal E} &=& \left\{\left[\fr{i}{2}\prt^jh_{j0}+a_0-a^j h_{j0}\right] + \left[-h_{j0}\right]p^j \right\} \nonumber \\ 
         &&  +\ga^0 \left\{-\half m h_{00} \right\} \nonumber \\ 
         &&  +\ga^0\ga^j\ga_5 \left\{\left[\fr{1}{4}\ve_{jkl}\prt^kh^{0l}-b_j+\half b^k \hbar_{jk}\right]\right\} 
 \quad , \text{ and} \nonumber \\ 
{\cal O} &=& \ga^5 \left\{ -b_0 + b^j h_{j0} \right\} \nonumber \\ 
         &&  + \ga^0\ga^j \left\{\left[\fr{i}{4}\prt^lh_{l0}+a_j-\half a^k\hbar_{jk}\right] 
                                 +\left[-\half\hbar_{jk}\right]p^k\right\} \quad . 
\eea 
In this expression, 
the perturbation terms have been sorted according to their status as $4\times 4$ gamma matrices: 
Terms in ${\cal E}$ have nonzero entries only in the upper-left and lower-right $2\times 2$ blocks, 
while terms in ${\cal O}$ have nonzero entries only in the upper-right and lower-left $2\times 2$ blocks. 
This sorting is useful for performing a Foldy-Wouthuysen transformation 
\cite{Foldy:1949wa,Kostelecky:1999zh,Lehnert:2004ri} to obtain 
a nonrelativistic hamiltonian 
that approximates the physics of \eqs{Hrel} and (\ref{PEO}) for low-energy fermions. 
This hamiltonian may then be used with conventional perturbation theory 
to derive experimental signals. 

\subsection{Predictions Prior to Explicit Calculation} 
\citelabel{estimates} 

Before performing explicit calculations, 
it is worth predicting the types of effects that may appear. 

\subsubsection{Dependence on $a_\mu$.} 
In the Minkowski-spacetime SME action, 
the coefficient $a_\mu$ for a single fermion 
may be removed by a field redefinition $\ps\rightarrow\exp{[if(a_\mu x^\mu)]}\ps$. 
In curved spacetime, however, 
where $a_\mu$ may depend on spacetime position, 
this field redefinition may only be used to remove one component, say, $a_0$. 
More precisely: If all four components $a_\mu$ for a particular spinor field $\ps$ are nonzero, 
then we may find a function $f$ 
such that the redefined spinor acts according to an action with $a_0=0$; 
however, the other three components $a_j$ for the redefined spinor generically will be nonzero and 
will depend on both the original $a_j$ and the original $a_0$. 

Since one component may be removed in an extended region 
(rather than at just a single point), 
all derivatives of this component may also be removed. 
Thus, rather than the 16 independent derivatives $\prt_\mu a_\nu$ that seem to exist, 
we expect all physically meaningful effects to depend on at most 12 independent derivatives. 

Moreover, 
when we perform a Foldy-Wouthuysen transformation to extract a nonrelativistic hamiltonian, 
the coefficient $a_\mu$ behaves like 
the electromagnetic potential $-qA_\mu$ in conventional physics. 
We therefore expect that measurable physical effects will depend 
at most on analogues of the field-strength components, 
$\prt_\mu a_\nu - \prt_\nu a_\mu$, 
and as part of the kinetic-energy-like term 
$\fr{1}{2m}(\vec{p}+\vec{a})^2$. 

It is worth briefly discussing the similarity of $a_\mu$ with $-qA_\mu$. 
They appear identically in the action for a Dirac fermion, 
and hence act identically for the calculations done in the current work. 
However, 
they are {\itshape not} the same, 
as a physical theory is not defined purely through its action. 
Other properties of a theory's constituents must be considered. 
In the case of $a_\mu$ versus $-qA_\mu$, 
it suffices to consider $U(1)$ transformations $\ps\rightarrow e^{i\theta}\ps$. 
The electromagnetic potential $-qA_\mu$ transforms as $-qA_\mu \rightarrow -qA_\mu+\prt_\mu \theta$, 
while the Lorentz-violation coefficients $a_\mu$ are invariant: $a_\mu \rightarrow a_\mu$.  

For the sake of completion, 
we preserve all components of $a_\mu$ in the explicit calculations that follow, 
though it will be seen that these predictions are vindicated. 

\subsubsection{Order-of-Magnitude Estimates of Sensitivities.} 
Before performing explicit calculations, 
it is worth estimating the size of terms that could appear in the nonrelativistic hamiltonian. 
We will then only explicitly calculate terms that are likely to either give relatively large effects 
or yield sensitivity to previously-unstudied combinations of Lorentz-violation coefficients. 

Let $k$ denote a generic SME coefficient
(either $a_\mu$ or $b_\mu$ in this work), 
$\prt$ a generic spacetime derivative, 
$p$ a generic fermion 3-momentum component, 
$m$ a generic fermion mass, 
and $h$ a generic component of $h_\mn$. 
In $\hbar=c=1$ units, 
$k$, $p$, and $\prt$ have dimensions of mass, 
while $h$ is dimensionless. 
For experiments involving atoms near Earth's surface, 
these factors have the approximate values shown in Table \ref{factormagnitude}. 
\begin{table} 
\begin{tabular}{c|c|c} 
 & Size for & Size for \\ 
Factor & nucleons & electrons \\ \hline 
$h$ & $10^{-9}$ & $10^{-9}$ \\ 
$p/m$ & $10^{-2}$ & $10^{-5}$ \\ 
$\prt h/m$ & $10^{-32}$ & $10^{-29}$ 
\end{tabular} 
\caption{Order-of-magnitude estimates of factors that may contribute to 
terms in the nonrelativistic hamiltonian.} 
\label{factormagnitude} 
\end{table} 

The Foldy-Wouthuysen transformation yields a nonrelativistic hamiltonian 
that is a sum of terms, 
each proportional to a power of $1/m^n$ for some positive integer $n$. 
(The rest-energy term $m$ is the lone exception.) 
Each term has a product of nonnegative powers of $k$, $\prt k$, $h$, $\prt h$, $p$, and $1/m$, 
with an appropriate number of factors to give the term an overall dimension of mass. 
As examples, terms like $a_0$, $a_j h^{j0}$, and $\de^{jm}\frac{\prt_j b_k}{m}\frac{p_m}{m}\si^k$ may appear. 

From Table \ref{factormagnitude}, 
it becomes clear that terms involving $\prt h$ are highly suppressed 
and do not yield useful sensitivities. 
We therefore neglect all terms involving derivatives $\prt_\al h_\mn$ for the rest of this work. 
Moreover, 
since the nonconstant nature of SME coefficients might be connected to the size of $h_{\mn}$, 
we neglect all derivatives of SME coefficients 
of second and higher order. 

\section{Explicit Calculation} 
\citelabel{explicit} 

In this section, 
the nonrelativistic $4\times 4$ hamiltonian $H_{NR}$ is calculated explicitly. 
The method for doing so is tedious but straightforward, 
as we can hijack the standard Foldy-Wouthuysen expressions that appear in textbooks \cite{bjorken1964relativistic} 
for, say, calculating the nonrelativistic hamiltonian 
for a fermion in the presence of an electromagnetic potential. 
Keeping terms up to order $1/m^2$, 
\begin{align} 
\label{fwgeneral} 
H_\text{NR} &= m\ga^0 + {\cal E} + 
 \fr{1}{4m}\ga^0 \left( 
   \{ {\cal P},{\cal P} \} + 2\{ {\cal P},{\cal O} \} + \{ {\cal O},{\cal O} \} 
   \right) 
 \nonumber \\ 
 &\quad -\fr{1}{8m^2} \left( 
   [{\cal P},[{\cal P},{\cal E}]] + [{\cal P},[{\cal O},{\cal E}]] + [{\cal O},[{\cal P},{\cal E}]] \right. \nonumber \\ 
   & \qquad\qquad \left. + i[{\cal P},\prt_0{\cal O}] + i[{\cal O},\prt_0{\cal O}] 
   \right) 
\quad . 
\end{align} 

The full result is unwieldy and difficult to interpret on its own. 
However, we may fruitfully compare the result to the Minkowski-spacetime, 
constant-SME-coefficient nonrelativistic hamiltonian $H_{\text{NR,Mink}}$ 
given by equation (24) of Ref.\ \cite{Kostelecky:1999zh}. 
We can then exploit analysis of $H_{\text{NR,Mink}}$ that has already been completed 
to aid our understanding of weakly-curved-spacetime Lorentz violation. 

$H_{\text{NR,Mink}}$ includes all fermion-associated Lorentz-violation coefficients $a_\mu,\ldots,H_\mn$. 
However, it suffices for this work to preserve only $a_\mu$, $b_\mu$, $c_\mn$, and $d_\mn$, 
setting $e_\mu$, $f_\mu$, $g_{\mu\nu\la}$, and $H_\mn$ to zero: 
\begin{align} 
\label{MinkHNR} 
H_\text{NR,Mink} 
 &= \ga^0 \left\{ 
   m +\fr{1}{2}\de^{mn}\fr{p_m p_n}{m} 
   \right\} 
   \nonumber \\ 
 &\quad +\openone \left\{ 
   [a_0] 
   +[m(c^{0m}+c^{m0})]\fr{p_m}{m} 
   \right\} 
   \nonumber \\ 
 &\quad +\ga^0 \left\{ 
   [-mc_{00}] 
   +[\de^{mk} a_k]\fr{p_m}{m} \right. \nonumber \\ 
   & \qquad\qquad +\left. [-mc^{mn}-\fr{1}{2}m\de^{mn}c_{00}]\fr{p_m p_n}{m^2} 
   \right\} 
   \nonumber \\ 
 &\quad +\ga^q\ga_5 \left\{ 
   [md_{q0}] 
   +[-b_0\de^m_q]\fr{p_m}{m} \right. \nonumber \\ 
   & \qquad\qquad\quad +\left. [-m\de^m_q d^{0n}-\fr{1}{2}m\de^m_q d^{n0}]\fr{p_m p_n}{m^2} 
   \right\} 
   \nonumber \\ 
 &\quad +\ga^0\ga^q\ga_5 \left\{ 
   [-b_q] 
   +[m\de^{mk}d_{qk}+m\de^m_q d_{00}]\fr{p_m}{m} \right. \nonumber \\ 
   &\qquad\qquad\qquad +\left.[\fr{1}{2}\de^{mn}b_q -\fr{1}{2}\de^m_q \de^{nk}b_k]\fr{p_m p_n}{m^2} 
   \right\} \quad . 
\end{align} 
(In this expression and all following analysis, 
$\openone$ denotes the $4\times 4$ identity matrix.) 

As an example of the comparison that can be done, 
$H_{\text{NR}}$ contains the term 
\beq 
H_{\text{NR}} \supset \openone \left\{ \left[ 
 -m h^{m0}
 -\frac{i\prt_{[j} b_{k]}}{4m} {\ve_p}^{mj} 
  \left(\de^{kp} - \hbar^{kp}\right) 
 \right] \frac{p_m}{m} 
 \right\} \quad . 
\label{exampleterm}
\eeq 
This involves the same operator as the term 
$\openone\left\{ mc^{0m} +mc^{m0} \right\}\frac{p_m}{m}$ 
in $H_{\text{NR,Mink}}$, 
implying that this derivative effect in weakly curved spacetime 
acts like an effective value of a Minkowski-space SME coefficient: 
\beq 
\left( mc^{0m}+mc^{m0} \right)_{\text{eff}} = 
 -m h^{m0}
 -\frac{i\prt_{[j} b_{k]}}{4m} {\ve_p}^{mj} \left( 
  \de^{kp} - \hbar^{kp} 
  \right) 
 \quad . 
\eeq 
The combination $mc^{0m}+mc^{m0}$ has already been analyzed and bounded in 
existing works \cite{Hohensee:2013cya,Altschul:2006pv,Wolf:2006uu,Kostelecky:2008ts}. 
For example, its value in a nonrotating Sun-centered frame 
$mc^{TX}+mc^{XT}$ associated with electrons is known to be 
smaller than about $10^{-18}$ GeV \cite{Hohensee:2013cya,Altschul:2010na}. 
We can exploit this result to estimate that 
$\abs{\prt_Y b_Z-\prt_Z b_Y} \alt 10^{-21}$ GeV$^2$ for electrons. 
Further analysis of this sort appears in Section \ref{analysis}. 

\begin{table*} 
\begin{displaymath} 
\begin{array}{c|c|l}  
\text{Operator} & \text{(Minkowski) effective coefficient} & \text{Weakly-curved-spacetime coefficient} \\ \hline\hline 
\openone & (a_0)_\text{eff} & 
  a_0 + a_k h^{k0} 
   \\ \hline 

\openone\frac{p_m}{m} & (mc^{0m}+mc^{m0})_\text{eff} & 
 m h^{m0} 
 -\frac{\prt_{[j} b_{k]}}{4m} \left( 
   {\ve_p}^{mj}\de^{kp} -{\ve_p}^{mj}\hbar^{kp} 
   \right) 
  \\ \hline 

\ga^0 & (-mc_{00})_\text{eff} & 
  -\frac{1}{2}m h_{00}
  -\frac{i\prt_{(j} a_{k)}}{2m} \left( 
    \de^{jk} -\hbar^{jk} +\fr{1}{2}\de^{jk}h_{00} 
    \right) 
   \\ \hline  

\ga^0\frac{p_m}{m} & (\de^{mk}a_k)_\text{eff} & 
  a_k \left( 
    \de^{km} -\hbar^{km} +\fr{1}{2}\de^{km} h_{00} 
  \right) 
   \\ \hline  

\ga^0\frac{p_m p_n}{m^2} & (-mc^{mn}-\fr{1}{2}m\de^{mn}c_{00})_\text{eff} & 
 -\fr{1}{2}m\left( \hbar^{mn} -\fr{1}{2}m\de^{mn}h_{00} \right) 
  \\ \hline  

\ga^q\ga_5 & (md_{q0})_\text{eff} & 
  \frac{\prt_{[j} a_{k]}}{2m} \left( 
    {\ve^{jk}}_q 
    -{\ve^{jp}}_q\hbar^k_p 
    +\fr{1}{2}{\ve^{jk}}_q h_{00}  
    \right) \\ 
 & & -\frac{i\prt_j b_0}{2m} \left( 
    -\de^j_q +\fr{1}{2}\hbar^j_q -\fr{1}{2}\de^j_q h_{00} 
    \right) \\ 
 & & +\frac{i\prt_{(j} b_{k)}}{2m} \left( \de^j_q h^{k0} \right) 
     +\frac{i\prt_{[j} b_{k]}}{2m} \left( \de^j_q h^{k0} \right) 
   \\ \hline  

\ga^q\ga_5\frac{p_m}{m} & (-b_0 \de^m_q)_\text{eff} & 
  b_0 \left( 
    -\de^m_q +\fr{1}{2}\hbar^m_q -\fr{1}{2}\de^m_q h_{00} 
    \right) 
  +b_k \left( 
    -\de^m_q h^{k0} 
    \right) 
   \\ \hline  

\ga^q\ga_5\frac{p_m p_n}{m^2} & (-m\de^m_q d^{0n}-\fr{1}{2}m\de^m_q d^{n0})_\text{eff} & 
  0 
   \\ \hline  

\ga^0\ga^q\ga_5 & (-b_q)_\text{eff} & 
  b_k \left( 
    -\de^k_q + \fr{1}{2}\hbar^k_q 
    \right) 
   \\ \hline  

\ga^0\ga^q\ga_5\frac{p_m}{m} & (m\de^{mk}d_{qk}+m\de^m_q d_{00})_\text{eff} & 
  \frac{\prt_{[j} a_{0]}}{2m} \left( 
    {\ve_q}^{jm}  + {\ve_q}^{p[j}{\hbar^{m]}}_p 
    \right) 
  +\frac{\prt_{[j} a_{k]}}{2m} \left( 
    {\ve^{jm}}_q h^{k0} 
    \right) 
    \\ 
  & & +\frac{i\prt_{(j} b_{k)}}{4m} \left( 
    {\ve^{pj}}_q{\ve^{km}}_p -\de^m_q\hbar^{jk} +\de^{mj}\hbar^k_q 
    \right) 
    \\ 
  & & -\frac{i\prt_{[j} b_{k]}}{4m} \left( 
    {\ve^{pm}}_q{\ve^{jk}}_p -{\ve^{pj}}_q{\ve^{km}}_p 
    +\de^j_q\hbar^{km} -\de^{mj}\hbar^k_q 
    \right) 
   \\ \hline  

\ga^0\ga^q\ga_5\frac{p_m p_n}{m^2} & (\fr{1}{2}\de^{mn}b_q - \fr{1}{2}\de^m_q\de^{nk} b_k)_\text{eff} & 
  \fr{1}{2}b_k \left( 
    {\ve_q}^{p(m}{\ve^{n)k}}_p +{\de_q}^{(m}\hbar^{n)k} 
    -\fr{1}{2}\de^{mn}\hbar^k_q -\fr{1}{2}\de^k_q\hbar^{mn} 
    \right) 
\end{array} 
\end{displaymath} 
\caption{Comparison of the nonrelativistic hamiltonians in Minkowski and weakly-curved spacetime. 
The Minkowski-spacetime hamiltonian may be regarded as 
an effective cartesian scalar product of the first and second columns 
(plus the conventional Minkowski-space hamiltonian), 
while the weakly-curved-spacetime hamiltonian may be regarded as 
an effective cartesian scalar product 
of the first and third columns 
(plus the conventional Minkowski-space hamiltonian).} 
\label{tableeffectiveparams} 
\end{table*} 

The full form of $H_\text{NR}$ is given by 
\begin{align} 
\label{HNR} 
H_\text{NR} 
 &= \ga^0 \left\{ 
   m +\fr{1}{2}\de^{mn}\fr{p_m p_n}{m} 
   \right\} 
   \nonumber \\ 
 &\quad +\openone \left\{ 
   [a_0]_\text{eff} 
   +[m(c^{0m}+c^{m0})]_\text{eff}\fr{p_m}{m} 
   \right\} 
   \nonumber \\ 
 &\quad +\ga^0 \left\{ 
   [-mc_{00}]_\text{eff} 
   +[\de^{mk} a_k]_\text{eff}\fr{p_m}{m} \right. \nonumber \\ 
   & \qquad\qquad \left. +[-mc^{mn}-\fr{1}{2}m\de^{mn}c_{00}]_\text{eff}\fr{p_m p_n}{m^2} 
   \right\} 
   \nonumber \\ 
 &\quad +\ga^q\ga_5 \left\{ 
   [md_{q0}]_\text{eff} 
   +[-b_0\de^m_q]_\text{eff}\fr{p_m}{m} \right. \nonumber \\ 
   & \qquad\qquad\quad \left. +[-m\de^m_q d^{0n}-\fr{1}{2}m\de^m_q d^{n0}]_\text{eff}\fr{p_m p_n}{m^2} 
   \right\} 
   \nonumber \\ 
 &\quad +\ga^0\ga^q\ga_5 \left\{ 
   [-b_q]_\text{eff} 
   +[m\de^{mk}d_{qk}+m\de^m_q d_{00}]_\text{eff}\fr{p_m}{m} \right. \nonumber \\ 
   & \qquad\qquad\qquad \left. +[\fr{1}{2}\de^{mn}b_q -\fr{1}{2}\de^m_q \de^{nk}b_k]_\text{eff}\fr{p_m p_n}{m^2} 
   \right\} 
 \quad , 
\end{align} 
where the full set of effective coefficients is collected in Table \ref{tableeffectiveparams}. 
It is worth reiterating a description of the two theories that are being compared in 
this Table. 
\begin{enumerate} 
\item The 2nd column of Table \ref{tableeffectiveparams} refers to the minimal SME for a free Dirac fermion 
 in Minkowski spacetime, 
 with nonzero and spacetime-constant coefficients $a_\mu$, $b_\mu$, $c_\mn$, and $d_\mn$. 
 The nonrelativistic $4\times 4$ hamiltonian for this theory is denoted $H_\text{NR,Mink}$, 
 and is given explicitly by \eq{MinkHNR}. 
 Alternatively, 
 $H_\text{NR,Mink}$ may be reconstituted by adding the 
 effective cartesian scalar product of the first and second columns of Table \ref{tableeffectiveparams} 
 to the conventional Minkowski-space hamiltonian. 
\item The 3rd column of Table \ref{tableeffectiveparams} refers to the minimal SME for a free Dirac fermion 
 in weakly curved spacetime, 
 with nonzero and spacetime-dependent coefficients $a_\mu$ and $b_\mu$. 
 The nonrelativistic $4\times 4$ hamiltonian for this theory is denoted $H_\text{NR}$. 
 This hamiltonian is not given explicitly in this work, 
 but may be easily reconstituted by plugging the relations defined in Table \ref{tableeffectiveparams} 
 into \eq{HNR}. 
 Alternatively, 
 $H_\text{NR}$ may be reconstituted by adding the 
 effective cartesian scalar product of the first and third columns of Table \ref{tableeffectiveparams} 
 to the conventional Minkowski-space hamiltonian. 
\end{enumerate} 

Each nonrelativistic hamiltonian describes both fermions (with the upper-left $2\times 2$ block) 
and antifermions (with the lower-right $2\times 2$ block). 
When applying either hamiltonian to most nonrelativistic systems, 
however, the antifermion portion is irrelevant. 
Extracting the fermion portion simply amounts to keeping only the upper-left $2\times 2$ block. 
The result of this extraction can be found most easily by replacing the operators 
$\openone$ and $\ga^0$ with the $2\times 2$ identity matrix, 
and replacing 
$\ga^q\ga_5$ and $\ga^0\ga^q\ga_5$ with $\si^q$.

\section{Analysis} 
\citelabel{analysis} 
In this section, 
we try to gain physical insight into $H_\text{NR}$. 
We first study the hermiticity of $H_\text{NR}$ 
and follow by isolating the dominant contribution to $H_\text{NR}$ 
from each derivative that appears. 
We then briefly consider the C, P, and T properties 
that can be associated with each combination of SME coefficients 
that appears therein. 
Finally, we determine the extent to which 
existing experiments are sensitive 
to each combination of derivatives. 

\subsection{Hermiticity and Dominant Terms} 
\citelabel{hermiticity} 

The Foldy-Wouthuysen process is a unitary transformation, 
which guarantees that the nonrelativistic hamiltonian is hermitian. 
However, this hermiticity is not obvious. 
For example, the term $\fr{1}{2}i \ga^q\ga_5 \prt_q b_0/m$ 
is clearly nonhermitian. 

The basic issue is that a product of two hermitian operators 
is itself hermitian if and only if the operators commute. 
In Minkowski spacetime, $b_0$ is independent of position, 
and therefore it commutes with each momentum operator $p_j$ 
(as does the matrix $\ga^q\ga_5$). 
As a result, the term $-\ga^q\ga_5 b_0 p_q/m$ is hermitian in Minkowski spacetime. 
In curved spacetime, however, $b_0$ depends on position, 
and therefore $[p_j,b_0]\ne 0$. 
As a result, $-\ga^q\ga_5 b_0 p_q/m$ is not hermitian in curved spacetime. 

While some individual terms may not be hermitian, 
they form combinations that are hermitian when added. 
For example, the combination 
\begin{displaymath} 
\ga^q\ga_5 \left( 
\frac{-b_0 p_q +\fr{1}{2}i\prt_q b_0}{m} 
\right) 
\end{displaymath} 
of the two examples given above 
is hermitian even though neither term is individually. 
Each other nonhermitian term in \eq{HNR} forms a combination with 
one or more other terms 
such that the result is hermitian. 

\begin{table*} \begin{displaymath} 
\begin{array}{c@{\hspace{0.5cm}}c@{\hspace{0.5cm}}c@{\hspace{0.5cm}}c} 
\text{Operator}     
  & \text{Minkowski-spacetime coefficient} 
  & \text{Weakly-curved-spacetime coefficient} 
  & \text{Intuitive equivalent} \\ \hline 
\openone\fr{p_m}{m} 
  & mc^{0m}+mc^{m0} 
  & -\fr{\prt_{[j} b_{k]}}{4m} {\ve^{mj}}_p \de^{kp} 
  & \text{curl}(\vec{b}) \\ 
\ga^0               
  & -mc_{00} 
  & -\fr{i\prt_{(j} a_{k)}}{2m} \de^{jk} 
  & i\, \text{div}(\vec{a}) \\ 
\ga^q\ga_5          
  & md_{q0} 
  & \fr{\prt_{[j} a_{k]}}{2m} {\ve^{jk}}_q 
   +\fr{i\prt q b_0}{2m} 
  & \text{curl}(\vec{a}) +i\, \text{grad}(b_0) \\ 
\ga^0\ga^q\ga_5\fr{p_m}{m} 
  & m\de^{mk}d_{qk}+m\de^m_q d_{00} 
  & \fr{1}{4m}{\ve^{jm}}_q \left( \prt_j a_0 -\prt_0 a_j \right) 
  & \text{grad}(a_0) -\prt_0\vec{a} \\ 
  & 
  & +\fr{i\prt_{(j}b_{k)}}{4m}{\ve^{pj}}_q {\ve^{km}}_p 
  & i\prt_{(j} b_{k)} \\ 
  & 
  &  -\fr{i\prt_{[j}b_{k]}}{4m} \left( {\ve^{pm}}_q {\ve^{jk}}_p +{\ve^{pj}}_q {\ve^{mk}}_p \right) 
  & i\, \text{curl}(\vec{b})
\end{array} 
\end{displaymath} 
\caption{Dominant appearance of each derivative. 
Note that $\prt_0 a_0$, $\prt_0 b_0$, and $\prt_0 b_k$ are absent, 
while only the trace part of $\prt_{(j}a_{k)}$ appears.} 
\label{dominant} 
\end{table*} 

Many of the terms in $H_\text{NR}$ are suppressed by factors of $h_\mn$ 
relative to other terms. 
Further, some terms (those containing only $m$, $p_m$, and/or $h_\mn$) are Lorentz symmetric 
while each of the non-derivative terms involving $a_\mu$ or $b_\mu$ 
has been studied elsewhere. 
It is therefore interesting to isolate the dominant term including each derivative 
of an SME coefficient. 
These dominant effects are summarized in Table \ref{dominant}. 

Note that three derivatives, 
$\prt_0 a_0$, $\prt_0 b_0$, and $\prt_0 b_k$, 
are entirely absent from $H_{\text{NR}}$. 
Moreover, the symmetric part of $\prt_j a_k$ only appears as a trace to leading order; 
the off-diagonal parts only appear when suppressed by $h^{jk}$. 
This echoes the appearance in conventional electrodynamics 
of $\prt_j A_k$ as part of $\fr{1}{2m}(\vec{p}-q\vec{A})^2$.

\subsection{C, P, and T Analysis.} 

As an aside, 
we may use the correspondence between 
the Minkowski and curved-spacetime hamiltonians 
to study the C, P, and T properties of interactions 
associated with SME coefficients. 
The interesting coefficients that appear in $H_{\text{NR}}$ 
are spacetime derivatives of $a_\nu$ and $b_\nu$, 
namely, $\prt_\mu a_\nu$ and $\prt_\mu b_\nu$. 
In the nonrelativistic approximation, though, 
space and time components are separated. 
In addition, coupling to $h_\mn$ may affect the CPT properties of some interactions. 

In Table \ref{cptproperties}, we summarize the C, P, and T properties of operators connected to 
SME coefficients for free Dirac fermions. 

\begin{table}[b] 
\begin{displaymath} 
\begin{array}{l|ccc|l} 
\text{Minkowski-Spacetime} & & & & \text{Weakly-Curved-Spacetime} \\ 
\text{Coefficients} & \text{C} & \text{P} & \text{T} & \text{Coefficients} \\ \hline 
c_{00},\; c_{jk} &+&+&+& h_{00},\; \hbar_{jk},\; i\prt_{(j} a_{k)} \\ 
b_k              &+&+&-& b_k \\  
b_0              &+&-&+& b_0,\; b_k h^{k0} \\  
c_{0j},\; c_{j0} &+&-&-& h^{j0},\; \prt_{[j} b_{k]} \\ 
a_0              &-&+&+& a_0,\; a_k h^{k0} \\ 
d_{j0},\; d_{0j} &-&+&-& \prt_{[j} a_{k]},\; i\prt_j b_0,\; i\prt_j b_k h^{k0} \\ 
d_{00},\; d_{jk} &-&-&+& \prt_{[j} a_{0]},\; \prt_{[j}a_{k]}h^{k0},\; i\prt_{(j}b_{k)},\; i\prt_{[j}b_{k]} \\
a_k              &-&-&-& a_k 
\end{array} 
\end{displaymath} 
\caption{C, P, and T properties associated with derivatives of SME coefficients.} 
\label{cptproperties} 
\end{table} 
From this table, 
the following rules can be extracted 
for relating  the C, P, and T properties 
of interactions associated directly with $a_\mu$ and $b_\mu$ 
to the C, P, and T properties of 
interactions associated with their derivatives. 
\begin{enumerate} 
\item Application of a time derivative leaves C and P unchanged, 
 but reverses T. 
\item Application of a spatial derivative leaves C and T unchanged, 
 but reverses P. 
\item Multiplication by $h_{00}$ or $h_{jk}$ leaves all C, P, and T properties unchanged. 
\item Multiplication by $h^{k0}$ leaves C unchanged, 
 but reverses both P and T. 
\item Multiplication by $i$ leaves P unchanged, 
 but reverses both C and T. 
\end{enumerate} 
These rules are summarized in Table \ref{cptrules}. 
\begin{table}[b] 
\begin{displaymath} 
\begin{array}{c|ccc} 
 & \multicolumn{3}{c}{\text{Effect on}} \\ 
\text{Factor} & \text{C} & \text{P} & \text{T} \\ \hline 
\prt_0                    & + & + & - \\ 
\prt_j                    & + & - & + \\ 
h_{00} \text{ or } h_{jk} & + & + & + \\ 
h^{k0}                    & + & - & - \\ 
i                         & - & + & - \\ 
\end{array} 
\end{displaymath} 
\caption{Rules for determining the C, P, and T properties 
of interactions associated with derivatives of $a_\mu$ and $b_\mu$.} 
\label{cptrules} 
\end{table}

\subsection{Sensitivity of Completed Experiments} 
\citelabel{sensitivity} 

While the Standard-Model Extension breaks particle Lorentz symmetry, 
it preserves observer symmetry. 
This means that its action takes the same form 
in every coordinate frame. 
If we restrict attention to frames where the metric 
can be written $g_{\mn}=\et_{\mn}+h_{\mn}$ with $\abs{h_{\mn}}\ll 0$, 
the action takes the form of \Eq{action}. 
If we further restrict attention to nonrelativistic systems 
such as slow-moving atoms and nuclei, 
all prior results of the current work hold, 
including the nonrelativistic hamiltonian (\ref{HNR}). 

Systems that are of interest in connecting the SME to experiment 
include a frame attached to the surface of Earth 
and the Sun-centered non-rotating frame \cite{Kostelecky:1999mr,Bluhm:2001rw,Bluhm:2003un,Kostelecky:2002hh} 
conventionally used for analysis of Lorentz violation. 
Sensitivity of experiments to the Minkowski-spacetime SME 
is typically expressed with respect to Sun-frame coordinates $(T,X,Y,Z)$. 

Evaluation of terms in Table \ref{tableeffectiveparams} 
requires taking the nonconstant nature of the coefficients into account. 
For example, calculating the expectation value of the term $-\ga^0\ga^q\ga_5 b_q$ 
in a state $\ket{\ps}$ 
involves an integral $\int d^3 x \abs{\ps(x)}^2 b_q(x)$. 
In general, 
the spatial dependence of $b_q$ will depend on the underlying theory, 
and so evaluation of this integral is model dependent. 
However, we may often make some progress by assuming that 
$b_q$ does not vary strongly, 
and thus may be approximated by its average value 
over a relevant spatial region. 
In fact, once we decide to take seriously 
the notion that Lorentz-violation coefficients may vary with position, 
we are forced to interpret all published sensitivities 
(such as those summarized in Ref.\ \cite{Kostelecky:2008ts}) in this or similar fashion. 
Many of the derivative terms may be treated in the same way without difficulty. 

Once we make this approximation, 
we can immediately apply existing bounds on Minkowski-spacetime coefficients 
to many coefficient derivatives. 
For example, it has been found that $\abs{\tb_X}\alt 10^{-33}$ GeV for the neutron \cite{Brown:2010dt}. 
In Minkowski spacetime, $md_{XT}$ contributes to $\tb_X$, 
and so $\fr{1}{2m}(\prt_Y a_Z - \prt_Z a_Y)$ contributes to it in curved spacetime. 
The experiment determining this limit occurred on Earth's surface, 
so it bounds the average value of $(\prt_Y a_Z - \prt_Z a_Y)$ over the volume 
of the solar system swept out by Earth during its orbit. 
We therefore find that the average value of $\abs{\prt_Y a_Z - \prt_Z a_Y} \alt 10^{-33}$ GeV$^2$ for neutrons. 
Several other bounds can be derived in similar fashion, 
and are listed as numbers without parentheses in Table \ref{boundtable}. 

Interpretation of the nonhermitian derivative terms is more complicated. 
Consider $\fr{1}{2}i \ga^q\ga_5 \prt_q b_0/m$ again as an example. 
If we try to simply approximate $\prt_q b_0$ to an average value, 
then its expectation value for an atomic state 
yields an imaginary energy shift. 
This cannot be the entire story. 
As described earlier, 
this term is partnered with $-\ga^q\ga_5 b_0 p_q/m$ 
to get a hermitian combination. 
To get a real number for the energy shift, 
we must take the nonconstant nature of $b_0$ seriously 
when evaluating the expectation value of the combined term 
$\ga^q\ga_5 \left( \frac{-b_0 p_q +\fr{1}{2}i\prt_q b_0}{m} \right)$. 
Within this expression, 
there is significant interplay between $\prt_q b_0$ 
and the nonconstant nature of $b_0$; 
that interplay, in fact, 
is critical in finding a real energy shift. 
Determining the exact nature of the interplay, 
however, 
is problematic, 
and dependent on the underlying model. 

Nevertheless, 
the expectation value of this combined term 
is likely to include order-one dependence on 
the average value of $\prt_q b_0/m$ over the relevant spatial region. 
Thus, we may make a rough but plausible estimate of 
the sensitivity of some experiments to $\prt_q b_0$. 
The results of this sort of analysis appear in Table \ref{boundtable} 
with parentheses 
to denote the stronger assumptions that must be made in deriving these estimates. 

\begin{table*} 
\begin{displaymath} 
\begin{array}{c||c||cc|cc|cc} 
\text{Weakly-curved-spacetime} & \text{Dominant sensitivity in terms}    & \multicolumn{6}{c}{\text{Sensitivity/GeV}^2\text{ and Reference}} \\ 
\text{coefficient} & \text{of Minkowski-space coefficient} & \multicolumn{2}{c|}{\text{Electron}} 
 & \multicolumn{2}{c|}{\text{Proton}} & \multicolumn{2}{c}{\text{Neutron}} \\ \hline 
\prt_{[X} a_{T]}   & 2m\abs{\tH_{XT}} & 10^{-29}   & \citemath{Heckel:2008hw}   
 & - &                                & 10^{-26}   & \citemath{Cane:2003wp} \\ 
\prt_{[Y} a_{T]}   & 2m\abs{\tH_{YT}} & 10^{-29}   & \citemath{Heckel:2008hw}   
 & - &                                & 10^{-26}   & \citemath{Cane:2003wp} \\ 
\prt_{[Z} a_{T]}   & 2m\abs{\tH_{ZT}} & 10^{-29}   & \citemath{Heckel:2008hw}   
 & - &                                & 10^{-27}   & \citemath{Cane:2003wp} \\ \hline 
\prt_{[Y} a_{Z]}   & m\abs{\tb_{X}}   & 10^{-34}   & \citemath{Heckel:2008hw}   
 & 10^{-33} & \citemath{Kornack:2008zz}   & 10^{-33}   & \citemath{Brown:2010dt}   \\ 
\prt_{[Z} a_{X]}   & m\abs{\tb_{Y}}   & 10^{-34}   & \citemath{Heckel:2008hw}  
 & 10^{-33} & \citemath{Kornack:2008zz}   & 10^{-33}   & \citemath{Brown:2010dt}   \\ 
\prt_{[X} a_{Y]}   & m\abs{\tb_{Z}}   & 10^{-32}   & \citemath{Heckel:2008hw}  
 & 10^{-28} & \citemath{Peck:2012pt}      & 10^{-29}   & \citemath{Brown:2010dt}   \\ 
\de^{JK}\prt_J a_K & 2m\abs{\tc_{TT}} & (10^{-21}) & \citemath{Altschul:2010na} 
 & (10^{-11}) & \citemath{Kostelecky:2010ze} & (10^{-11}) & \citemath{Kostelecky:2010ze} \\ \hline 
\prt_X b_T         & 2m\abs{\tb_{X}}  & (10^{-34}) & \citemath{Heckel:2008hw}  
 & (10^{-33}) & \citemath{Kornack:2008zz} & (10^{-33}) & \citemath{Brown:2010dt} \\ 
\prt_Y b_T         & 2m\abs{\tb_{Y}}  & (10^{-34}) & \citemath{Heckel:2008hw}  
 & (10^{-33}) & \citemath{Kornack:2008zz} & (10^{-33}) & \citemath{Brown:2010dt} \\ 
\prt_Z b_T         & 2m\abs{\tb_{Z}}  & (10^{-32}) & \citemath{Heckel:2008hw}  
 & (10^{-28}) & \citemath{Peck:2012pt}    & (10^{-29}) & \citemath{Peck:2012pt}  \\ \hline
\prt_{(Y} b_{Z)}   & 2m\abs{\td_{YZ}} & (10^{-29}) & \citemath{Heckel:2008hw}  
 & -  &                               & (10^{-26}) & \citemath{Cane:2003wp} \\ 
\prt_{(Y} b_{Z)}   & 4m\abs{\tH_{XT}} & (10^{-29}) & \citemath{Heckel:2008hw}  
 & -  &                               & (10^{-26}) & \citemath{Cane:2003wp} \\ 
\prt_{(Z} b_{X)}   & 2m\abs{\td_{ZX}} & (10^{-29}) & \citemath{Heckel:2008hw}  
 & -  &                               & -          \\ 
\prt_{(Z} b_{X)}   & 4m\abs{\tH_{YT}} & (10^{-29}) & \citemath{Heckel:2008hw}  
 & -  &                               & (10^{-26}) & \citemath{Cane:2003wp} \\ 
\prt_{(X} b_{Y)}   & 2m\abs{\td_{XY}} & (10^{-29}) & \citemath{Heckel:2008hw}  
 & -  &                               & (10^{-27}) & \citemath{Cane:2003wp} \\ 
\prt_{(X} b_{Y)}   & 4m\abs{\tH_{ZT}} & (10^{-29}) & \citemath{Heckel:2008hw}  
 & -  &                               & (10^{-27}) & \citemath{Cane:2003wp} \\ \hline 
\prt_{[Y} b_{Z]}   & 2m\abs{\tc_{TX}} & 10^{-21}   & \citemath{Hohensee:2013cya,Altschul:2006pv} 
 & 10^{-20} & \citemath{Wolf:2006uu}      & 10^{-5}    & \citemath{Kostelecky:2010ze} \\ 
\prt_{[Y} b_{Z]}   & 2m\abs{\tH_{XT}} & (10^{-29}) & \citemath{Heckel:2008hw}   
 & -  &                               & (10^{-26}) & \citemath{Cane:2003wp} \\ 
\prt_{[Z} b_{X]}   & 2m\abs{\tc_{TY}} & 10^{-21}   & \citemath{Hohensee:2013cya,Altschul:2006pv} 
 & 10^{-20} & \citemath{Wolf:2006uu}      & 10^{-5}    & \citemath{Kostelecky:2010ze}    \\ 
\prt_{[Z} b_{X]}   & 2m\abs{\tH_{YT}} & (10^{-29}) & \citemath{Heckel:2008hw}  
 & -  &                               & (10^{-26}) & \citemath{Cane:2003wp} \\ 
\prt_{[X} b_{Y]}   & 2m\abs{\tc_{TZ}} & 10^{-23}   & \citemath{Hohensee:2013cya,Altschul:2006pv} 
 & 10^{-20} & \citemath{Wolf:2006uu}      & 10^{-5}    & \citemath{Kostelecky:2010ze}    \\ 
\prt_{[X} b_{Y]}   & 2m\abs{\tH_{ZT}} & (10^{-29}) & \citemath{Heckel:2008hw}  
 & -  &                               & (10^{-27}) & \citemath{Cane:2003wp} \\ 
\end{array} 
\end{displaymath} 
\caption{Maximal sensitivity to derivatives of SME coefficients 
from already-completed experiments. 
Sensitivities written with parentheses require further assumptions than 
those written without parentheses.} 
\label{boundtable} 
\end{table*} 

\section{Summary}
\citelabel{Summary} 
This work has studied the variance of SME Lorentz-violation coefficients 
with spacetime position. 
Such variation is likely to be necessary in curved spacetime. 
We have calculated the nonrelativistic hamiltonian that may be used for determining 
physical consequences of varying $a_\mu$ and $b_\mu$ coefficients. 
We found that nontrivial but solvable issues with hermiticity arise, 
and presented the C, P, and T properties of derivative-associated operators. 
Finally, we have found the maximal sensitivity of completed experiments 
to variation of SME coefficients. 



\begin{acknowledgments} 
This work was supported in part by Berry College 
and the Indiana University Center for Spacetime Symmetries (IUCSS). 
\end{acknowledgments}

%

\end{document}